\documentclass[12pt]{article}
\usepackage{amsfonts,amssymb,amsmath}
\usepackage[dvips]{epsfig}
\begin{document}
\title{Excited Coherent States Attached to Landau Levels }
\author{B. Mojaveri\thanks{Email: bmojaveri@azaruniv.ac.ir}\hspace{2mm} and \hspace{2mm} A. Dehghani\thanks{Email: a\_dehghani@tabrizu.ac.ir,  alireza.dehghani@gmail.com}\\
{\small {\em Department of Physics, Azarbaijan Shahid Madani
University, PO Box 51745-406, Tabriz, Iran\,}}\\
{\small {\em Department of Physics, Payame Noor University, P.O.Box
19395-3697 Tehran, I.R. of Iran. \,}}} \maketitle
\begin{abstract}
A new scheme is proposed to design excited coherent states, $|\beta,
\alpha; \mathbf{n}\rangle:={{{a^{\dag}}^{\mathbf{n}}}|\beta,
\alpha\rangle}$. Where the states $|\beta, \alpha\rangle$ denote the
Glauber two variable minimum uncertainty coherent states, which
minimize minimum uncertainty conditions while carrier nonclassical
properties too and $\mathbf{n}$ is an integer. They are converted
into the Agarwal's type of the photon added coherent states,
arbitrary Fock states and the Glauber two variable coherent states,
respectively depending on which of the parameters $\alpha, \beta$
and $\mathbf{n}$ equal to zero. It has been shown that the
resolution of identity condition is realized with respect to an
appropriate measure on the complex plane, too. We have compared our
results with the similar quantum states of Agarwal's type and seen
that in our case amount of quantum fluctuations are much
controllable. Moreover, we have also more flexibility to establish
and set-out of their features. Also, there is a discussion on the
statistical properties, can unveil (non-)classical properties of
these states. For instance their Poissonian statistics are
significant similar to what we saw earlier in the Glauber two
variable coherent states. Interestingly, depending on the particular
choice of the parameters of the above scenarios, we are able to
determine the status of compliance with squeezing properties in
field quadratures. The last stage is devoted to some theoretical
framework to generate them in cavities. 

\end{abstract}

\section{Introduction}
In 1991, Agarwal et al. \cite{Agarwal} introduced photon added
coherent states (PACSs) $|\alpha, \mathbf{n}\rangle$ which are
obtained from the excited states of the Schr\"{o}dinger's
non-spreading wave packets, $|\alpha\rangle$, that minimize the
uncertainty in the measurement of the position as well as momentum
operators and follow the classical motion. In other word, the states
$|\alpha, \mathbf{n}\rangle$ are emerged through an iterated action
of creation operator $a^{\dag}$ on the coherent states
$|\alpha\rangle$, i.e.
\begin{eqnarray}
&&\hspace{-15mm}|\alpha, \mathbf{n}\rangle: \equiv
{{a^{\dag}}^{\mathbf{n}}}|\alpha\rangle,\hspace{4mm}{\mathbf{n}}\in
N_{0},\end{eqnarray}where $a^{\dag}$ refer to the creation operator
of a simple harmonic oscillator. They are intermediate between a
single-photon Fock state (fully quantum-mechanical)
$|\mathbf{n}\rangle$ and a coherent (classical) one
$|\alpha\rangle$, these states offer the opportunity to closely
follow the smooth transition between the particle-like and the
wavelike behavior of light. In other word they reduces to two or
more distinguishably different states in different limits, as
\begin{eqnarray}&&\hspace{-20mm}\alpha=0\hspace{23mm}\hspace{5mm}\mathbf{n}=0\nonumber\\
&&\hspace{-35mm}|\mathbf{n}\rangle\hspace{10mm}\Longleftarrow\hspace{10mm}|\alpha,
\mathbf{n}\rangle\hspace{10mm}\Longrightarrow\hspace{5mm}|\alpha\rangle\nonumber.\end{eqnarray}
Their mathematical and physical properties were studied in details,
for instance they exhibit phase squeezing and sub-Poissonian
statistics of the field. Also, an interesting theoretical framework
was proposed by many authors about how such states can be generated
in nonlinear processes in cavities. Fortunately, their aspiration
becomes a reality and in 2004 Zavatta et al. \cite{Zavatta} set out
the experimental generation of single-photon-added coherent states
and their complete characterization by quantum tomography. Dynamical
squeezing of these states and their classification in a special
class of non-linear coherent states were done by many authors
\cite{Dodonov, Sivakumar}. The idea of construction similar states,
in correspondence with discrete and unitary representation of Lie
algebras $SU(1, 1)$, extended by one of the authors \cite{dehghani}.

The coherent states for a charged particle in a magnetic field were
first introduced by I. A. Malkin et. al 1968 \cite{manko}, also the
relevant developments can be found in papers \cite{manko1, manko2,
manko3}. By using of unitary displacement operators corresponding to
the Weyl-Heisenberg algebras (Glauber unitary displacement operator)
acting on the standard Schrodinger coherent states, Feldman et al
1970 \cite{Feldman} introduced the Glauber two-variable minimum
uncertainty coherent states $|\beta, \alpha\rangle$ for the problem
of an electron in the constant magnetic field. These states minimize
the Heisenberg uncertainty relation between the position and
momentum operators. Recently, In Ref. \cite{dehghani0}, we have
presented another minimum uncertainty coherent states for the Landau
levels based on the action of unitary displacement operators
associated to the Weyl-Heisenberg algebras on the Klauder-Perelomov
coherent states of $su(1; 1)$ and $su(2)$ algebras, too. One can be
pointed to their outstanding non-classical properties such as
quadrature squeezing and anti-bunching effects, however they
minimize the uncertainty condition too, which is very
important in this respect and distinguishes them from other minimum uncertainty states \cite{dehghani1}.\\
Therefore, based on the fact that the Glauber two-variable minimum
uncertainty coherent states $|\beta, \alpha\rangle$ could be good
alternative instead the standard Schr\"{o}dinger's coherent states
\cite{Feldman, dehghani0, dehghani1}. Then our main motivation on
this document is to construct new kinds of two variable photon-added
coherent states $|\beta, \alpha; \mathbf{n}\rangle$ in terms of the
states $|\beta, \alpha\rangle$.  So the introduced photon added
coherent states $|\beta, \alpha; \mathbf{n}\rangle$ are not a
trivial generalization of the well known photon added coherent
states $|\alpha; \mathbf{n}\rangle$. As motioned in the abstract
section, investigation on their statistical properties highlight the
considerable features rather than the well known photon added
coherent state $|\alpha; \mathbf{n}\rangle$. Hence, in section 3, we
have established a new class of photon-added states $|\beta, \alpha;
{\mathbf{n}}\rangle$ in correspondence with the two-dimensional
Landau levels. They can be converted into different status that we
summarize them in the following diagram
\begin{eqnarray}&&\hspace{-6mm}|\beta, {\mathbf{n}}\rangle\nonumber\\
&&\hspace{-12mm}\alpha=0\Uparrow\hspace{13mm}\nonumber\\
&&\hspace{-42mm}\vspace{15mm}\nonumber\\
&&\hspace{-50mm}|{\mathbf{n}},
-{\mathbf{n}}\rangle\hspace{10mm}\Longleftarrow
\hspace{10mm}|\beta, \alpha; {\mathbf{n}}\rangle\hspace{10mm}\Longrightarrow\hspace{10mm}|\beta, \alpha\rangle\nonumber,\\
&&\hspace{-33mm}\alpha=\beta=0\hspace{29mm}\mathbf{n}=0\nonumber\\
&&\hspace{-38mm}\vspace{15mm}\nonumber\\
&&\hspace{-12.5mm}\beta=0\hspace{2mm}{{\Downarrow}}\nonumber\\
&&\hspace{-6mm}|\alpha, \mathbf{n}\rangle\nonumber\end{eqnarray}
where $|\beta, \alpha\rangle$, $|\mathbf{n}, -\mathbf{n}\rangle$,
$|\alpha, {\mathbf{n}}\rangle$ and $|\beta, {\mathbf{n}}\rangle$
denote the well known Glauber two-variable minimum uncertainty
coherent states, Fock number states as Landau levels with lowest
$z$-angular momentum and the two different copies of the Agarwal's
type of the photon add coherent states attached to the two different
modes, $a$ and $b$, of the simple harmonic oscillators,
respectively. In order to realize the resolution of the identity, we
have found the positive definite measures on the complex plane in
sub.section 3.2. Also, sub.section 3.4 is devoted to discuss about
the fact that these states can be considered as an eigenstate of
certain annihilation operators, and be interpreted as nonlinear
coherent states with special nonlinearity functions. Furthermore, it
has been discussed in detail, in sub.sections 3.5 and 3.6, that they
have indeed nonclassical features such as squeezing, anti-bunching
effects and sub-Poissonian statistics, too. Finally, in section 4,
we propose an approach of generation of the states $|\beta,
\alpha\rangle$ as well as $|\beta, \alpha; \mathbf{n}\rangle$, which
can be produced by interaction of a two level atom with two mode
cavity fields.

\section{Reviews on Landau levels, the Weyl-Heisenberg algebras and their coherency}
In Refs. \cite{ Feldman, dehghani0}, it has been shown that the
symmetric-gauge Landau Hamiltonian corresponding to the motion of an
electron on a flat surface in the presence of an unified magnetic
field in the positive direction of z axis, given by
\begin{eqnarray}
&&\hspace{-1.5cm}
H=\hbar\omega\left({a}^{\dag}a+\frac{1}{2}\right)=\hbar\omega\left({b}^{\dag}b+\frac{1}{2}\right)-\omega
L_{3},\end{eqnarray} with $L_{3} =
-i\frac{\partial}{\partial\varphi}$. It has an infinite-fold
degeneracy on the Landau levels, that is
\begin{eqnarray}
&&\hspace{-1.5cm} H |n,m\rangle= \hbar\omega\left(n +
\frac{1}{2}\right) |n,m\rangle,\end{eqnarray} in which Landau
cyclotron frequency is expressed in terms of the value of the
electron charge, its mass, the magnetic field strength $B_{ext}$ and
also the velocity of light as $\omega= \frac{eB_{ext}}{Mc}$. Here,
$m$ is an integer number and $n$ is a nonnegative one together with
$n\geq -m$ limitation. Each pair of operators $({a}, {a}^{\dag})$
and $({b}, {b}^{\dag})$ have the following explicit forms in terms
of the polar coordinates $0 < r <\infty, 0\leq \varphi< 2\pi$ for
two-dimensional flat surface,
\begin{eqnarray}
&&\hspace{-1.5cm}a= -e^{i\varphi}
\sqrt{\frac{\hbar}{2M\omega}}\left(\frac{\partial}{\partial
r}+\frac{i}{r}\frac{\partial}{\partial
\varphi}+\frac{M\omega}{2\hbar}r\right),\hspace{4mm} {a}^{\dag}=
e^{-i\varphi}
\sqrt{\frac{\hbar}{2M\omega}}\left(\frac{\partial}{\partial
r}-\frac{i}{r}\frac{\partial}{\partial
\varphi}-\frac{M\omega}{2\hbar}r\right),\\
&&\hspace{-1.5cm}b= e^{-i\varphi}
\sqrt{\frac{\hbar}{2M\omega}}\left(\frac{\partial}{\partial
r}-\frac{i}{r}\frac{\partial}{\partial
\varphi}+\frac{M\omega}{2\hbar}r\right),\hspace{4mm} {b}^{\dag}=
-e^{i\varphi}
\sqrt{\frac{\hbar}{2M\omega}}\left(\frac{\partial}{\partial
r}+\frac{i}{r}\frac{\partial}{\partial
\varphi}-\frac{M\omega}{2\hbar}r\right),\end{eqnarray} and form two
separate copies of Weyl-Heisenberg algebra,
\begin{eqnarray}
&&\hspace{-1.5cm}[a, {a}^{\dag}] = 1, [b, b^{\dag}] = 1, [a,
b^{\dag}] = [a^{\dag}, b] = [a, b] = [a^{\dag}, b^{\dag}]
=0,\end{eqnarray} with the unitary representations as
\begin{eqnarray}
&&\hspace{-1.5cm} a |n,m\rangle = \sqrt{n} |n - 1,m + 1\rangle
,\hspace{4mm}
a^{\dag} |n - 1,m + 1\rangle = \sqrt{n} |n,m\rangle ,\\
&&\hspace{-1.5cm} b |n,m\rangle = \sqrt{n + m} |n,m - 1\rangle
,\hspace{4mm} b^{\dag} |n,m -1\rangle = \sqrt{n + m}
|n,m\rangle.\end{eqnarray} Landau levels are orthonormal with
respect to integration over the entire plane, that is
\begin{eqnarray}
&&\hspace{-1.5cm}\langle n, m|n',
m'\rangle:=\int^{2\pi}_{\varphi=0}\int^{\infty}_{r=0}{\langle r,
\varphi|n, m\rangle^{\ast}\langle r, \varphi|n', m'\rangle
rdrd\varphi}=\delta_{nn'}\delta_{mm'},\end{eqnarray} in which
\begin{eqnarray}
&&\hspace{-1.5cm}\langle r, \varphi|n,
m\rangle=\sqrt{\frac{n!}{\pi(n+m)!}\left(\frac{M\omega}{2\hbar}\right)^{m+1}}
r^{m}e^{im\varphi}e^{-\frac{M\omega}{4\hbar}r^2}L^{(m)}_{n}\left({M\omega
r^2}/{2\hbar}\right),\end{eqnarray} is the polar coordinate
representation of Landau levels in terms of the associated Laguerre
functions.

The normalized Weyl- Heisenberg coherent states corresponding to one
of the classes mentioned above is constructed as:
\begin{eqnarray}&&\hspace{-7mm}|\alpha\rangle^{b}_{n} = e^{\alpha b^{\dag} -
\overline{\alpha} b} |n, -n\rangle =
e^{-\frac{|\alpha|^2}{2}}\sum^{\infty}_{m=-n}{\frac{\alpha^{n+m}}{\sqrt{(n+m)!}}|n,
m\rangle},\\
&&\hspace{-1.5cm}\langle r, \varphi|\alpha\rangle^{b}_{n}
=\sqrt{\frac{M\omega}{2\pi
\hbar}}\frac{\left(\alpha-\sqrt{\frac{M\omega}{2\hbar}}r
e^{-i\varphi}\right)^{n}}{\sqrt{n!}}e^{-\frac{|\alpha|^2}{2}+\alpha\sqrt{\frac{M\omega}{2\hbar}}r
e^{i \varphi}-\frac{M\omega}{4\hbar}r^2},\end{eqnarray} in which
$\alpha$ is an arbitrary complex variable (with the polar form
$\alpha = Re^{i\phi}; 0 \leq R <\infty, 0 \leq \phi <2\pi$). It is
an infinite superposition of degenerate levels, and also satisfies
the following eigenvalue equations
\begin{eqnarray}&&\hspace{-5mm}b | \alpha\rangle^{b}_{n} = \alpha|
\alpha\rangle^{b}_{n},\\
&&\hspace{-7mm}H|\alpha\rangle^{b}_{n} = \hbar\omega(n + 1/2)
|\alpha\rangle^{b}_{n}.\end{eqnarray} It is well known that the
resolution of the identities on entire complex plane are realized
for such coherent states by the measure $d\mu(\alpha) =
\frac{1}{\pi}RdRd\phi$. Clearly, one can deduce that coherent states
$|\alpha\rangle^{b}_{n}$ form complete as well as orthonormal bases,
i.e.
\begin{eqnarray}^{b}_{n}\langle
\alpha|\alpha\rangle^{b}_{n'}=
\delta_{nn'},\hspace{5mm}\sum^{\infty}_{n=0}|\alpha\rangle^{b}_{n}\hspace{0.5mm}{_n^{b}}\langle\alpha|=I^{Obl.}_{0}=\sum^{\infty}_{n=0}|n,-n\rangle\langle
n,-n|,\end{eqnarray} then, it is obvious that the quantization
described by the coherent states $|\alpha\rangle^{b}_{n}$ can be
used to the representation of the ladder operators $a$ and
$a^{\dag}$,
\begin{eqnarray}&&\hspace{-7mm}a|\alpha\rangle^{b}_{n}=\sqrt{n}|\alpha\rangle^{b}_{n-1},\hspace{5mm}
a^{\dag}|\alpha\rangle^{b}_{n-1}=\sqrt{n}|\alpha\rangle^{b}_{n},\end{eqnarray}
and allow us to consider separately coherency property for the
coherent states $\left|\alpha\right\rangle_{n}^{b}$. The recent
coherency follows by a new complex variable $\beta$ involved in
Glauber unitary displacement operator corresponding to the
Weyl-Heisenberg algebras:
\begin{eqnarray}
&&\hspace{-10mm}\left|\beta,\alpha\right\rangle=e^{\beta
a^{\dagger}-\bar{\beta}a}\left|\alpha\right\rangle_0^{b}=
e^{-|\beta|^2/2}\sum_{n=0}^{\infty}\frac{\beta^n}{\sqrt{n!}}\left|\alpha\right\rangle_n^{b},\\
&&\hspace{-20mm}\langle r,
\varphi\left|\beta,\alpha\right\rangle=\sqrt{\frac{M\omega}{2\pi\hbar}}
e^{-(|\alpha|^2+|\beta|^2)/2+\alpha\beta+\sqrt{\frac{M\omega}{2\hbar}}\,r
(\alpha e^{i\varphi}-\beta
e^{-i\varphi})-\frac{M\omega}{4\hbar}r^2}.
\end{eqnarray}
It is worth to mention that eigenvalue equations corresponding to
this two-variable coherent state is
$a\left|\beta,\alpha\right\rangle=\beta\left|\beta,\alpha\right\rangle$
and
$b\left|\beta,\alpha\right\rangle=\alpha\left|\beta,\alpha\right\rangle$(
for some notational details on such two-variable coherent states see
Refs. \cite{ Feldman, dehghani0, dehghani1}).
\section{Generalized Photon Added Coherent States By Levels as the Glauber Two Variable Coherent States And Their Properties }
We introduce the state $|\beta, \alpha; {\mathbf{n}}\rangle$ defined
by
\begin{eqnarray}&&\hspace{-28mm}|\beta,
\alpha;
{\mathbf{n}}\rangle:=\frac{{a^{\dag}}^{{\mathbf{n}}}}{\sqrt{\langle\beta,
\alpha|a^{{\mathbf{n}}}{{a^{\dag}}^{{\mathbf{n}}}}|\beta,
\alpha\rangle}}|\beta,
\alpha\rangle\nonumber\\
&&\hspace{-10mm}=\frac{e^{-\frac{|\beta|^2}{2}}}{\sqrt{\langle\beta,
\alpha|a^{{\mathbf{n}}}{{a^{\dag}}^{{\mathbf{n}}}}|\beta,
\alpha\rangle}}\sum^{\infty}_{n=0}{\beta^{n}\sqrt{\frac{({\mathbf{n}}+n)!}{n!^2}}|\alpha\rangle^{b}_{n+{\mathbf{n}}}},\hspace{4mm}{\mathbf{n}}\in
\mathbb{N},\end{eqnarray} here, the factor
$\frac{1}{\sqrt{\langle\beta,
\alpha|a^{{\mathbf{n}}}{{a^{\dag}}^{{\mathbf{n}}}}|\beta,
\alpha\rangle}}$ is prepared so that $|\beta, \alpha;
{\mathbf{n}}\rangle$ is normalized, i.e. $\langle\beta, \alpha;
{\mathbf{n}}|\beta, \alpha; {\mathbf{n}}\rangle = 1$. Then we have
\begin{eqnarray}&&\hspace{-15mm}{{\langle\beta, \alpha|a^{{\mathbf{n}}}{{a^{\dag}}^{{\mathbf{n}}}}|\beta,
\alpha\rangle}}={\mathbf{n}}!L_{{\mathbf{n}}}(-|\beta|^2).\end{eqnarray}
Due to the orthogonality relation of Fock space basis (9), it
follows that overlapping of two different kinds of these normalized
states must be nonorthogonal in the following sense
\begin{eqnarray}&&\hspace{-15mm}{{\langle\beta, \alpha; {\mathbf{n}}'|\beta,
\alpha;
{\mathbf{n}}\rangle}}={\overline{\beta}}^{{\mathbf{n}}-{\mathbf{n}}'}\frac{{\mathbf{n}}'!L^{{\mathbf{n}}-{\mathbf{n}}'}_{{\mathbf{n}}'}(-|\beta|^2)}
{\sqrt{{\mathbf{n}}!L_{{\mathbf{n}}}(-|\beta|^2){\mathbf{n}}'!L_{{\mathbf{n}}'}(-|\beta|^2)}},\end{eqnarray}
which would be advantageous in the calculation of the expectation
values of observables.
\subsection{{\em{\bf{Coordinate Representation of $|\beta, \alpha;
{\mathbf{n}}\rangle$}}}} Based on a the Eqs. (19) and (20) also
using by the spatial representation of the Glauber two variable
coherent states $|\beta, \alpha\rangle$, Eq. (18), one can calculate
to be taken as
\begin{eqnarray}&&\hspace{-30mm}\langle r, \varphi|\beta,
\alpha;
{\mathbf{n}}\rangle=\frac{e^{-\frac{|\alpha|^2+|\beta|^2}{2}+\alpha\beta}}{\sqrt{\pi{\mathbf{n}}!L_{{\mathbf{n}}}(-|\beta|^2)}2^{\frac{{\mathbf{n}+1}}{2}}\left(\frac{M\omega}{
\hbar}\right)^\frac{{\mathbf{n}-1}}{2}}\nonumber\\
&&\hspace{-1mm}\times\left[e^{-i\varphi}\left(\frac{\partial}{\partial
r}-\frac{i}{r}\frac{\partial}{\partial\varphi}-\frac{M\omega}{2\hbar}r\right)\right]^{{\mathbf{n}}}e^{\sqrt{\frac{M\omega}{2\hbar}}r(\alpha
e^{i\varphi}-\beta
e^{-i\varphi})-\frac{M\omega}{4\hbar}r^2}.\end{eqnarray} For
instance we have
\begin{eqnarray}&&\hspace{-15mm}\langle r, \varphi|\beta,
\alpha; 1\rangle=\frac{\alpha
-\sqrt{\frac{M\omega}{2\hbar}}re^{-i\varphi}}{\sqrt{1+|\beta|^2}}\langle
r, \varphi|\beta, \alpha\rangle,\end{eqnarray}and so on.
\subsection{\bf{\em{Resolution of unity (or completeness)}}}
From equation (19) we see that the state $|\beta, \alpha;
{\mathbf{n}}\rangle$ is a linear combination of all number coherent
states starting with $n = \mathbf{n}$. In otherwords, the first
$\mathbf{n}$ number coherent states $n = 0, 1, . . ., \mathbf{n} -
1$, are absent from these states. Then, the unity operator in this
space is to be written as
\begin{eqnarray}&&\hspace{-20mm}I_{\mathbf{n}}=\sum_{n=0}^{\infty}{|\alpha\rangle^{b}_{n+\mathbf{n}}\hspace{1mm}_{n+\mathbf{n}}^{b}\langle
\alpha|}=
\sum_{n=\mathbf{n}}^{\infty}{|\alpha\rangle^{b}_n\hspace{1mm}^{b}_{n}\langle
\alpha|}=\sum_{n=\mathbf{n}}^{\infty}{|n, -n\rangle\langle n, -n|}
.\end{eqnarray} Evidently, in the right-hand side of the above
equation, the identity operator on the full Hilbert space does not
appear, because of the initial $\mathbf{n}$ states of the basis set
vectors are omitted.
\begin{figure}
\begin{center}
\epsfig{figure=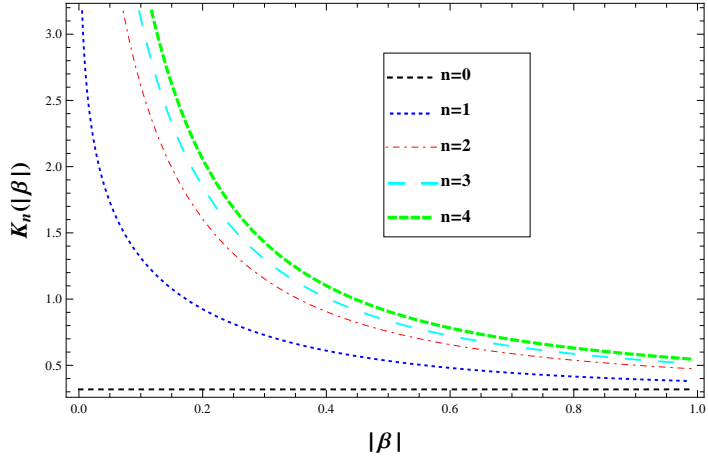,width=10cm}
\end{center}
\caption{\footnotesize Plots of the non-oscillating and positive
definite measures ${K}_{\textbf{n}}(|\beta|)$ in terms of $|\beta|$
for different values of $\mathbf{n}$. The dotted and horizontal
curve corresponds to  the Glauber two-variable minimum uncertainty
coherent states $|\beta, \alpha\rangle$.}
\end{figure}
This leads to the following resolution of unity via bounded,
positive definite and non-oscillating measures, $d\eta
_{\mathbf{n}}(|\beta|) := K_{\mathbf{n}}(|\beta|)
\frac{d|\beta|^2}{2}d\varphi$, in terms of the Meijer{'}s
G-function( see $\frac{7-811}{4}$ in \cite{Gradshteyn})
\begin{eqnarray}&&\hspace{-25mm}\oint_{\mathbb{C}(\beta)}{|\beta, \alpha; \mathbf{n}\rangle\hspace{1mm}
\langle \beta, \alpha; \mathbf{n}|}d\eta_{\mathbf{n}}(|\beta|)=I_{\mathbf{n}},\\
&&\hspace{-25mm}{K}_{\mathbf{n}}(|\beta|)=\frac{\mathbf{n}!}{\pi}e^{|\beta|^2}
L_{\mathbf{n}}\left(-|\beta|^2\right)G\left([[\hspace{1mm}],[\mathbf{n}]],[[0,
0],[\hspace{1mm}]],|\beta|^2\right).\end{eqnarray} We have plotted
the changes of the function ${K}_{\mathbf{n}}(|\beta|)$ in Figure 1.
\subsection{{\bf\em{Time Evolution }}}
As we discussed above, displaced number states
$\left|\alpha\right\rangle_n^{b}$ form complete set of orthonormal
states which satisfy an eigenvalue equation
\begin{eqnarray}
&&\hspace{-14mm}H\left|\alpha\right\rangle_n^{b}=\hbar\omega\left(n+\frac{1}{2}\right)\left|\alpha\right\rangle_n^{b}.\end{eqnarray}
Then, by acting the time evolution operator on the states (19) it
evolves in time as
\begin{eqnarray}
&&\hspace{-14mm}e^{-i\frac{t}{\hbar}H}|\beta, \alpha;
{\mathbf{n}}\rangle=\frac{e^{-\frac{|\beta|^2}{2}-i(\mathbf{n}+\frac{1}{2})\omega
t}}{\sqrt{\mathbf{n}!L_{\mathbf{n}}\left(-|\beta|^2\right)}}
\sum^{\infty}_{n=0}{\left(\beta e^{-i\omega
t}\right)^{n}\sqrt{\frac{({\mathbf{n}}+n)!}{n!^2}}|\alpha\rangle^{b}_{n+{\mathbf{n}}}}\nonumber\\
&&\hspace{12mm}=e^{-i(\mathbf{n}+\frac{1}{2})\omega t}|\beta
e^{-i\omega t}, \alpha; {\mathbf{n}}\rangle,\end{eqnarray} the
temporal stability follows easily, which illustrates the fact that
the time evolution of such states remain within the family of
coherent states.
\subsection{\bf{\em{Construction of Nonlinearity function}}}
In this section we construct the explicit form of the operator
valued of nonlinearity function associated to these new photon added
coherent states. Since, the coherent states $|\beta, \alpha\rangle$
satisfy the following eigenvalue equation
\begin{eqnarray}&&\hspace{-20mm}a|\beta, \alpha\rangle=\beta|\beta, \alpha\rangle,\nonumber\end{eqnarray}
so, multiplying both sides of this equation by
$\left(a^{\dag}\right)^{\mathbf{n}}$ yields
\begin{eqnarray}&&\hspace{-20mm}\left(a^{\dag}\right)^{\mathbf{n}}a|\beta, \alpha\rangle=\beta\left(a^{\dag}\right)^{\mathbf{n}}|\beta, \alpha\rangle\nonumber.\end{eqnarray}
Which, making use of the commutation relations $[a, a^{\dag}]=1$ and
the identities
\begin{eqnarray}&&\hspace{-20mm}\left(a^{\dag}\right)^{\mathbf{n}}a=a\left(a^{\dag}\right)^{\mathbf{n}}-\mathbf{n}\left(a^{\dag}\right)^{\mathbf{n}-1},\nonumber\end{eqnarray}
it leads to
\begin{eqnarray}&&\hspace{-18mm}
\left(1-\frac{\mathbf{n}}{a^{\dag}a+1}\right)a|\beta, \alpha;
\mathbf{n}\rangle=\beta|\beta, \alpha;
\mathbf{n}\rangle,\end{eqnarray} and pretends them as nonlinear
coherent states by the expression for nonlinearity function, in
terms of the number operator ${N}(=a^{\dag}a)$, as
\begin{eqnarray}&&\hspace{-20mm}1-\frac{\mathbf{n}}{N+1}.\end{eqnarray}
Obviously, it transforms to the identity operator for
$\mathbf{n}=0$.
\subsection{\bf{\em{Photon Number Distribution}}}
Another aspect of these states is revealed by considering their
photon number distribution $p^{\mathbf{n}}_{n, m}(\beta, \alpha)$,
which is the probability of finding an oscillator described by the
coherent state $|\beta, \alpha; \mathbf{n}\rangle$ in its $(n,
m)^{th}$ state
\begin{eqnarray} &&\hspace{-28mm}p^{\mathbf{n}}_{n, m}(\beta, \alpha)=
\frac{n!}{\mathbf{n}!(n+m)!(n-\mathbf{n})!^{2}}
\frac{|\alpha|^{2(n+m)}|\beta|^{2(n-\mathbf{n})}}{L_{\mathbf{n}}\left(-|\beta|^2\right)}e^{-|\alpha|^2-|\beta|^2}\nonumber\\
&&\hspace{-1cm}=\left[
\frac{n!}{\mathbf{n}!(n-\mathbf{n})!L_{\mathbf{n}}\left(-|\beta|^2\right)}
\frac{|\beta|^{2(n-\mathbf{n})}e^{-|\beta|^2}}{(n-\mathbf{n})!}\right]
\left[\frac{|\alpha|^{2(n+m)}}{(n+m)!}e^{-|\alpha|^2}\right],\hspace{5mm}n\geq\mathbf{n}.\end{eqnarray}
Clearly, depends on the choice of the parameters, we will see
different features from Poissonian to no-Poissonian distribution. In
fact, the dependence of the function $p^{\mathbf{n}}_{n, m}(\beta,
\alpha)$ on the parameters $\mathbf{n}, n, m$ and $\beta, \alpha$
provides more options in order to determine and especially control
of the expected distribution. In
a few cases the further analysis will focus on the latter issue.\\\\
$\diamondsuit$ Case $\mathbf{n}=0$:\\ Because of the fact that the
Eq. (19) dealing with the well known Glauber two variable coherent
states $|\beta, \alpha\rangle$ for $\mathbf{n}=0$ and it can be
considered as the states describe two coupled simple harmonic
oscillators, too. So, the probability of finding the state $|\beta,
\alpha; \mathbf{n}=0\rangle$ in its $(n, m)^{th}$ states follow
Poissonian regime, as we expected, i.e.
\begin{eqnarray} &&\hspace{-35mm}p^{0}_{n, m}(\beta, \alpha)=
\left[\frac{|\beta|^{2n}e^{-|\beta|^2}}{n!}\right]
\left[\frac{|\alpha|^{2(n+m)}}{(n+m)!}e^{-|\alpha|^2}\right].\end{eqnarray}
$\diamondsuit$ Case $n=0$ (Lowest Landau Level):\\
It is really interesting to note that, the probability of finding
the system in its lowest state obeys Poissonian distribution, i.e.
\begin{eqnarray} &&\hspace{-35mm}p^{\mathbf{n}}_{0, m}(\beta, \alpha)=
e^{-|\beta|^2}
\left[\frac{|\alpha|^{2m}}{m!}e^{-|\alpha|^2}\right]\end{eqnarray}
\\
$\diamondsuit$ Case $\mathbf{n}=n$:\\
Using the Eq. (31):
\begin{eqnarray} &&\hspace{-35mm}p^{\mathbf{n}=n}_{n, m}(\beta, \alpha)=
\left[\frac{e^{-|\beta|^2}}{L_{{n}}\left(-|\beta|^2\right)}\right]
\left[\frac{|\alpha|^{2(n+m)}}{(n+m)!}e^{-|\alpha|^2}\right],\end{eqnarray}
it indicates that the system carrier  a no-Poissonian distribution ,
except for $|\beta|=0$.\\\\
$\diamondsuit$ Case $n+m=0$ (Landau levels with lowest
$z$-angular momentum):\\
Another situation is related to the case $n+m=0$, when the above
mentioned system is in Landau levels with lowest $z$-angular
momentum. Our calculations show that it experiences a no-Poissonian,
i.e.
\begin{eqnarray} &&\hspace{-35mm}p^{\mathbf{n}}_{n, -n}(\beta, \alpha)=e^{-|\alpha|^2}
\left[
\frac{n!}{\mathbf{n}!(n-\mathbf{n})!L_{\mathbf{n}}\left(-|\beta|^2\right)}
\frac{|\beta|^{2(n-\mathbf{n})}e^{-|\beta|^2}}{(n-\mathbf{n})!}\right]\end{eqnarray}
\subsubsection{\em{\bf{Sub-Poissonian Statistics For The Field
In PACSs }}} As we know, a measure of the variance of the photon
number distribution is given by Mandel's $Q_{\mathbf{n}}(|\beta|)$
parameter \cite{Mandel}
\begin{eqnarray}
&&\hspace{-14mm}Q_{\mathbf{n}}(|\beta|)=
\frac{\langle{{N}}^2\rangle_{\mathbf{n}}-{\langle{{N}\rangle}_{\mathbf{n}}}^{2}
}{{\langle{{N}\rangle}_{\mathbf{n}}}}-1,\end{eqnarray} it provides a
convenient way of studying the nonclassical properties of the field.
For this reason we begin to calculate the expectation values of the
number operator ${N}$ and it's square in the basis of the PACSs
\begin{eqnarray}
&&\hspace{-20mm}{\langle{{N}\rangle}_{\mathbf{n}}}=(\mathbf{n}+1)\frac{L_{\mathbf{n}+1}(-|\beta|^2)}{L_{\mathbf{n}}(-|\beta|^2)}-1,\nonumber\\
&&\hspace{-20mm}\langle{{N}}^2\rangle_{\mathbf{n}}=(\mathbf{n}+1)(\mathbf{n}+2)\frac{L_{\mathbf{n}+2}(-|\beta|^2)}{L_{\mathbf{n}}(-|\beta|^2)}
-3(\mathbf{n}+1)\frac{L_{\mathbf{n}+1}(-|\beta|^2)}{L_{\mathbf{n}}(-|\beta|^2)}+1,\nonumber\end{eqnarray}
\begin{figure}
\begin{center}
\epsfig{figure=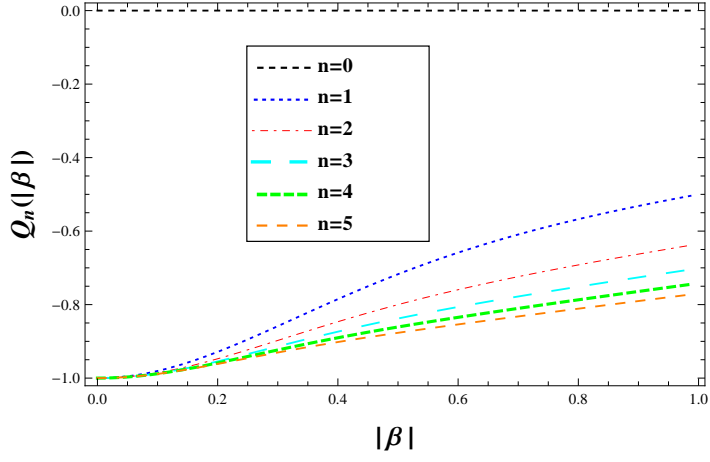,width=10cm}
\end{center}
\caption{\footnotesize Mandel's parameters
${Q}_{\mathbf{n}}(|\beta|)$ in the PACSs, versus $|\beta|$ for
different values of $\mathbf{n}$. The dotted and horizontal curve
$\left( {Q}_{\mathbf{n}=0}(|\beta|)=0\right)$ corresponds to the
standard Glauber coherent states.}
\end{figure}
Because of the structure of this function as illustrated in Figure
2, the states $|\beta, \alpha; \mathbf{n}\rangle$ show
sub-Poissonian statistics( or fully anti-bunching effects).
\subsection{\bf{\em{Minimum Uncertainty Condition }}}
Similar to what we have shown in our recent work \cite{dehghani0},
here we are looking for design ideas in the states $|\beta, \alpha;
\mathbf{n}\rangle$. From these, the uncertainty condition for the
field quadrature operators $x$ and it's conjugate momenta $p_{x}$
\begin{eqnarray} &&\hspace{-1.5cm}x =
\sqrt{\frac{\hbar}{2M\omega}}\left( b + b^{\dag} - a -
a^{\dag}\right),\hspace{4mm} p_{x} = \frac{i}{
2}\sqrt{\frac{M\hbar\omega}{2}} (b^{\dag} - b + a -
a^{\dag}),\end{eqnarray}follow
\begin{eqnarray}
&&\hspace{-14mm}\Delta:={\sigma^{(\mathbf{n})}_{xx}}{\sigma^{(\mathbf{n})}_{p_{x}p_{x}}}
-{\sigma^{(\mathbf{n})}_{xp_{x}}}^2,\end{eqnarray} where
${\sigma^{(\mathbf{n})}_{xx'}}\left( =\frac{1}{2}\langle{
xx'+xx'}\rangle_{\mathbf{n}}-\langle x\rangle_{\mathbf{n}}\langle
x'\rangle_{\mathbf{n}}\right)$ and the angular brackets denote
averaging over PACSs for which the mean values are well defined,
i.e.
\begin{eqnarray}
&&\hspace{-14mm}\langle x\rangle_{\mathbf{n}}={\langle\beta, \alpha;
\mathbf{n}|}x|\beta, \alpha;
\mathbf{n}\rangle.\nonumber\end{eqnarray} For instance, we have the
following expectation values
\begin{eqnarray}&&\hspace{-12mm}\left\langle b\right\rangle_{\mathbf{n}}
=\overline{\left\langle b^{\dag}\right\rangle}_{\mathbf{n}}=\alpha,\nonumber\\
&&\hspace{-12mm}\left\langle b^2\right\rangle_{\mathbf{n}}
=\overline{\left\langle \left(b^{\dag}\right)^{2}\right\rangle}_{\mathbf{n}}=\alpha^{2},\nonumber\\
&&\hspace{-12mm}\left\langle b^{\dag}b\right\rangle_{\mathbf{n}}
=\overline{\left\langle bb^{\dag}\right\rangle}_{\mathbf{n}}-1=|\alpha|^{2},\nonumber\\
&&\hspace{-12mm}\left\langle
a\right\rangle_{\mathbf{n}}=\overline{\left\langle
a^{\dag}\right\rangle}_{\mathbf{n}}=
\frac{L^{1}_{\mathbf{n}}\left(-|\beta|^2\right)}{L_{\mathbf{n}}\left(-|\beta|^2\right)}\beta,\nonumber\\
&&\hspace{-12mm}\left\langle
a^{2}\right\rangle_{\mathbf{n}}=\overline{\left\langle
\left({a^{\dag}}\right)^{2}\right\rangle}_{\mathbf{n}}
=\frac{L^{2}_{\mathbf{n}}\left(-|\beta|^2\right)}{L_{\mathbf{n}}\left(-|\beta|^2\right)}\beta^{2},\nonumber\\
&&\hspace{-12mm}\left\langle
a^{\dag}a\right\rangle_{\mathbf{n}}=\left\langle
aa^{\dag}\right\rangle_{\mathbf{n}}-1=\left\langle
N\right\rangle.\nonumber\end{eqnarray}Where they result
\begin{eqnarray}&&\hspace{-12mm}{\sigma^{(\mathbf{n})}_{xx}}
=\frac{\hbar}{M\omega}\left[\frac{(\mathbf{n}+1)L_{\mathbf{n}+1}\left(-|\beta|^2\right)+
|\beta|^2\cos(2\theta)L^{2}_{\mathbf{n}}\left(-|\beta|^2\right)-2|\beta|^2{\cos(\theta)}^{2}
\frac{L^{1}_{\mathbf{n}}\left(-|\beta|^2\right)^2}{L_{\mathbf{n}}\left(-|\beta|^2\right)}}
{L_{\mathbf{n}}\left(-|\beta|^2\right)}\right],\\
&&\hspace{-12mm}{\sigma^{(\mathbf{n})}_{p_{x}p_{x}}}
=\frac{M\hbar\omega}{4}\left[\frac{(\mathbf{n}+1)L_{\mathbf{n}+1}\left(-|\beta|^2\right)-
|\beta|^2\cos(2\theta)L^{2}_{\mathbf{n}}\left(-|\beta|^2\right)-2|\beta|^2{\sin(\theta)}^{2}
\frac{L^{1}_{\mathbf{n}}\left(-|\beta|^2\right)^2}{L_{\mathbf{n}}\left(-|\beta|^2\right)}}
{L_{\mathbf{n}}\left(-|\beta|^2\right)}\right],\\
&&\hspace{-12mm}{\sigma^{(\mathbf{n})}_{xp_{x}}}=\frac{\hbar}{2}
|\beta|^2\sin(2\theta)\left[\frac{L^{2}_{\mathbf{n}}\left(-|\beta|^2\right)-\frac{L^{1}_{\mathbf{n}}\left(-|\beta|^2\right)^2}{L_{\mathbf{n}}\left(-|\beta|^2\right)}}
{L_{\mathbf{n}}\left(-|\beta|^2\right)}\right].\end{eqnarray}
Clearly the minimum uncertainty condition is violated, except for
the values $\mathbf{n}=0$ or $|\beta|\gg 1$, which implies the
existence of non-classical properties that will be investigated.
\subsubsection{\bf{\em{Squeezing Properties}}} Our final step is to
reveal that measurements on the states $|\beta, \alpha;
\mathbf{n}\rangle$ come with squeezing for the field quadrature
operators $x$ and $p_{x}$. In accordance with Walls \cite{Walls}, a
set of quantum states are called squeezed states if they have less
uncertainty in one quadrature ($x$ or $p_{x}$) than coherent states.
Using the relations (39) and (40), it is easy to see that
${\sigma^{(\mathbf{n})}_{xx}}$ and
${\sigma^{(\mathbf{n})}_{p_{x}p_{x}}}$ are dependent on the complex
variable $\beta( =|\beta|e^{i\theta})$, the deformation parameter
$\mathbf{n}$ and the variations of the cyclotron frequency $\omega$
which comes from the variations of magnetic field $B_{ext}$ and
indicate that the squeezing properties could be varied by the
changes of the external magnetic field. Our calculations show that
these states exhibit squeezing in one of $x$ or $p_{x}$ quadratures,
clearly in different domains. In Figure 3(a), we have presented
${\sigma^{(\mathbf{n})}_{p_{x}p_{x}}}$ against $|\beta|$ for $
\theta= 0$ and different values of $\mathbf{n}(= 0, 1, 2, 3, 4, 5)$.
They become smaller than $\frac{1}{2}$ for sufficiently large
$|\beta|$. Although for the values $\mathbf{n} = 0, 1$ we will see
the fully squeezing in the whole range of $|\beta|$ and $\theta$,
i.e.
\begin{eqnarray}
&&\hspace{-14mm}\forall\hspace{1mm}|\beta|\in \mathbb{R}^{+} ,
\forall\hspace{1mm}\theta \in [0,
2\pi]\mid\hspace{2mm}{\sigma^{(\mathbf{n}=0, 1)}_{p_{x}p_{x}}}<
\frac{1}{2}.\nonumber\end{eqnarray}
\begin{figure}
\epsfig{figure=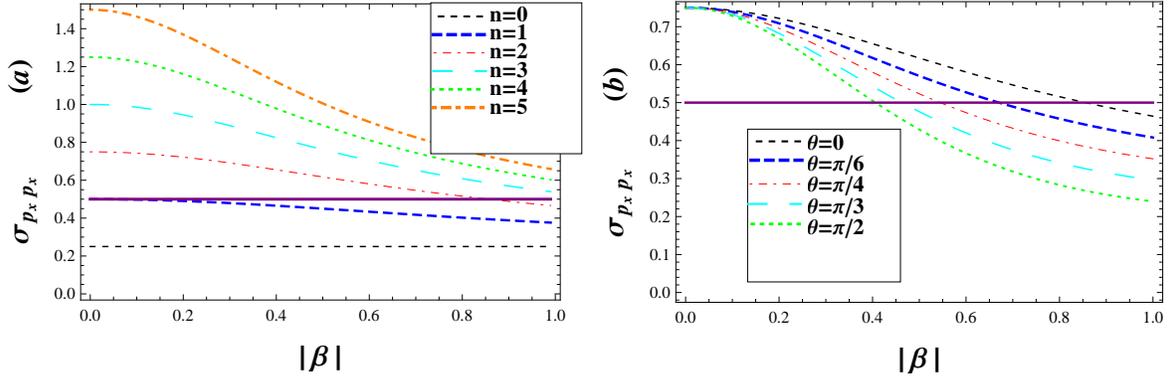,width=22cm} \caption{\footnotesize (a)
Squeezing in the $p_{x}$ quadrature against $|\beta|$ for $\theta=
0$ and different values of $\mathbf{n}= 0, 1, 2, 3, 4, 5$. (b)
Squeezing in the $p_{x}$ quadrature against $|\beta|$ for different
values of $\theta= 0, \pi/6, \pi/4, \pi/3, \pi/2$ where we fixed
$\textbf{n}=2$. The horizontal and solid line ${\sigma_{pp}}=
\frac{1}{2}$, is correspond to the canonical coherent state.}
\end{figure}
Another situation depicting the variation of
${\sigma^{(m)}_{p_{x}p_{x}}}$ in terms of $|\beta|$ for different
values of $\theta= 0, \frac{\pi}{6}, \frac{\pi}{4}, \frac{\pi}{3},
\frac{\pi}{2}$, while for fixed $\mathbf{n}=2$. Figure 3 (b) show
that, by decreasing $|\beta|$, the degree of squeezing is enhanced,
specially it passes the greatest value when $\theta$ reaches
$\frac{\pi}{2}$. It is worth to mention that, this feature is not
observed in the position coordinate $x$ for any value of the
parameters $\beta$ and $\mathbf{n}$. It is important to note that we
will see the fully squeezing in $|\beta, \alpha,
\mathbf{n}=0\rangle$ \cite{dehghani1}( the lowest horizontal line in
the left part of Figure 3(a)), this mean that
$\sigma_{p_xp_x}=\frac{M\hbar\omega}{4}<\frac{1}{2}$, while it
minimize the minimum uncertainty condition too ( i.e.
$\Delta=\sigma_{xx}
\sigma_{p_xp_x}-\sigma_{xp_x}^2=\frac{\hbar^2}{4}$)
\cite{dehghani0}. In other word, in addition to the requirement to
minimize the minimum uncertainty condition they carrier squeezing
properties too, which is unique in its kind.\\
\section{Theoretical Frameworks of Generation of the States $|\beta, \alpha\rangle$ and $|\beta, \alpha; \mathbf{n}\rangle$} Now, we propose theoretical frameworks to
generate these states in practice. We first discuss the way of
generating the two-variable coherent states, $|\beta,
\alpha\rangle$. Then the way of producing the "photon-added" states
$|\beta, \alpha; \mathbf{n}\rangle$ relevant to the two-variable
coherent states for the system has been reviewed in details.
\subsection{Generation of the States $|\beta, \alpha\rangle$} We consider a Hamiltonian in which a two level atom
interacts with two different modes of the cavity fields via an
intensity-dependent coupling also an external classical field. Based
on the rotating-wave approximation theory and the resonance
condition, the Hamiltonian becomes the following form (assuming
$\hbar = 1)$:
\begin{eqnarray}
&&\hspace{-20mm}
H=g(\sigma_{-}e^{i\varphi}+\sigma_{+}e^{i\varphi})+\Omega_1(a^{\dagger}\sigma_{-}+a\sigma_{+})+\Omega_1(b^{\dagger}\sigma_{-}+b\sigma_{+})
\end{eqnarray}
where two different copies of the creation and annihilation
operators $(a,a^{\dagger})$ and $(b,b^{\dagger})$, respectively,
describe the two different modes of the cavity fields. The
parameters $\Omega_{1}, \Omega_{2}$ and $g$ are referred to the
coupling coefficients of a two level atom with cavity and classical
fields, where the latter include the phase $\varphi$. Also, the
operators $\sigma_{-}=|g \rangle\langle e|$ and $\sigma_{+}=|e
\rangle\langle g|$ denote the atomic lowering and raising operators,
respectively, in terms of a two level atomic ground( $|g\rangle$)
and exited( $|e\rangle$) states. Here, we assume that the coupling
of the atom with the classical field be stronger than the coupling
of the atom with the cavity fields, i.e. $(g\gg \Omega_{1},
\Omega_{2})$.

Considering a situation in which the cavity is initially prepared in
the vacuum of the fields and, also, the atom in a superposition of
excited and ground states with equal weights:
\begin{eqnarray}
&&\hspace{-20mm}
\psi(0)=\left(\frac{|g\rangle+|e\rangle}{\sqrt{2}}\right)|0,
0\rangle .\end{eqnarray} Along with the time evolution operator
\cite{Zou}
\begin{eqnarray}
&&\hspace{-20mm} U(t)=R^{\dagger}T^{\dagger}(t)U_{eff}T(0)R,
\end{eqnarray}
where
\begin{eqnarray}
&&\hspace{-20mm} R=
e^{\frac{\pi}{4}(\sigma_{+}-\sigma_{-})}e^{\frac{i\varphi}{2}\sigma_{z}}
=\frac{1}{\sqrt{2}}(\hat{I}+\sigma_{+}-\sigma_{-})e^{\frac{i\varphi}{2}\sigma_{z}},\nonumber\\
&&\hspace{-20mm}T(t)=e^{ig\sigma_{z}t},\nonumber\\
&&\hspace{-20mm}U_{eff}=e^{\frac{i\Omega_{1}t}{2}(a^{\dagger}e^{-i\varphi}+ae^{i\varphi})\sigma_{z}}
e^{\frac{i\Omega_{2}t}{2}(b^{\dagger}e^{-i\varphi}+be^{i\varphi})\sigma_{z}}
\end{eqnarray}
, one can show that the system evolves into
{\footnotesize\begin{eqnarray}
&&\hspace{-5mm}\psi(t)=U(t)\psi(0)=R^{\dagger}T^{\dagger}(t)U_{eff}T(0)R\psi(0)\nonumber\\
&&\hspace{-4mm}=R^{\dagger}T^{\dagger}(t)\left\{\cos\left(\frac{\varphi}{2}\right)
e^{\frac{i\Omega_{1}t}{2}(a^{\dagger}e^{-i\varphi}+ae^{i\varphi})}
\,e^{\frac{i\Omega_{2}t}{2}(b^{\dagger}e^{-i\varphi}+be^{i\varphi})}|0,0\rangle|e\rangle\right.\nonumber\\
&&\hspace{12mm}\left.-i\sin\left(\frac{\varphi}{2}\right)\,e^{-\frac{i\Omega_{1}t}{2}(a^{\dagger}e^{-i\varphi}+ae^{i\varphi})}
\,e^{-\frac{i\Omega_{2}t}{2}(b^{\dagger}e^{-i\varphi}+be^{i\varphi})}|0,0\rangle|g\rangle\right\}\nonumber\\
&&\hspace{-4mm}=R^{\dagger}T^{\dagger}(t)\left\{\cos\left(\frac{\varphi}{2}\right)\,
e^{\frac{i\Omega_{1}t}{2}(a^{\dagger}e^{-i\varphi}+ae^{i\varphi})}
\left|\frac{-i\Omega_{2}t}{2}e^{-i\varphi}\right\rangle_{0}^{b}|e\rangle-i\sin\left(\frac{\varphi}{2}\right)\,e^{-\frac{i\Omega_{1}t}{2}(a^{\dagger}e^{-i\varphi}+ae^{i\varphi})}
\left|\frac{i\Omega_{2}t}{2}e^{-i\varphi}\right\rangle_{0}^{b}|g\rangle\right\}\nonumber\\
&&\hspace{-4mm}=R^{\dagger}T^{\dagger}(t)\left\{\cos\left(\frac{\varphi}{2}\right)
\left|\frac{i\Omega_{1}t}{2}e^{-i\varphi},\frac{-i\Omega_{2}t}{2}e^{-i\varphi}\right\rangle|e\rangle
-i\sin\left(\frac{\varphi}{2}\right)\left|\frac{-i\Omega_{1}t}{2}e^{-i\varphi},\frac{i\Omega_{2}t}{2}e^{-i\varphi}\right\rangle|g\rangle
\right\}\nonumber\\
&&\hspace{-4mm}=R^{\dagger}\left\{e^{-igt}\cos\left(\frac{\varphi}{2}\right)\,
\left|\frac{i\Omega_{1}t}{2}e^{-i\varphi},\frac{-i\Omega_{2}t}{2}e^{-i\varphi}\right\rangle|e\rangle
-ie^{igt}\sin\left(\frac{\varphi}{2}\right)\left|\frac{-i\Omega_{1}t}{2}e^{-i\varphi},\frac{i\Omega_{2}t}{2}e^{-i\varphi}\right\rangle|g\rangle\right\}\nonumber\\
&&\hspace{-4mm}=\frac{1}{\sqrt{2}}\left\{e^{-i(gt+\frac{\varphi}{2})}\cos\left(\frac{\varphi}{2}\right)\,
\left|\frac{i\Omega_{1}t}{2}e^{-i\varphi},\frac{-i\Omega_{2}t}{2}e^{-i\varphi}\right\rangle
-ie^{i(gt+\frac{\varphi}{2})}\sin\left(\frac{\varphi}{2}\right)|\frac{-i\Omega_{1}t}{2}e^{-i\varphi},\frac{i\Omega_{2}t}{2}
e^{-i\varphi}\rangle\right\}|g\rangle\nonumber\\
&&\hspace{-3mm}+\frac{1}{\sqrt{2}}\left\{e^{-i(gt+\frac{\varphi}{2})}\cos\left(\frac{\varphi}{2}\right)\,
\left|\frac{i\Omega_{1}t}{2}e^{-i\varphi},\frac{-i\Omega_{2}t}{2}e^{-i\varphi}\right\rangle
+ie^{i(gt+\frac{\varphi}{2})}\sin\left(\frac{\varphi}{2}\right)\left|\frac{-i\Omega_{1}t}{2}e^{-i\varphi},\frac{i\Omega_{2}t}{2}e^{-i\varphi}\right\rangle\right\}|e\rangle.
\end{eqnarray}}
By setting the parameters
$\alpha:=-\frac{i\Omega_{2}t}{2}e^{-i\varphi}$ and
$\beta:=\frac{i\Omega_{1}t}{2}e^{-i\varphi}$, the final state
$|\psi(t)\rangle$ becomes a general superposition of two variable
coherent states of Landau levels:
\begin{eqnarray}
&&\hspace{-17mm}|\psi(t)\rangle=\frac{1}{\sqrt{2}}\left\{e^{-i(gt+\frac{\varphi}{2})}\cos\left(\frac{\varphi}{2}\right)\,
|\beta,\alpha\rangle
-ie^{i(gt+\frac{\varphi}{2})}\sin\left(\frac{\varphi}{2}\right)|-\beta,-\alpha\rangle\right\}|g\rangle\nonumber\\
&&\hspace{-5mm}+\frac{1}{\sqrt{2}}\left\{e^{-i(gt+\frac{\varphi}{2})}\cos\left(\frac{\varphi}{2}\right)\,
|\beta,\alpha\rangle
+ie^{i(gt+\frac{\varphi}{2})}\sin\left(\frac{\varphi}{2}\right)|-\beta,-\alpha\rangle\right\}|e\rangle.
\end{eqnarray}
Then, depending on the state in which the atom is prepared,
different situations of the quantized cavity fields can be obtained.
For instance, if the atom is prepared in excited or ground state,
the fields will become, respectively, into:
\begin{eqnarray}
&&\hspace{-5mm}\psi_{g}(t)=N_{g}\left\{e^{-i\frac{\varphi}{2}}\cos\left(\frac{\varphi}{2}\right)\,
|\beta,\alpha\rangle-ie^{i\frac{\varphi}{2}}\sin\left(\frac{\varphi}{2}\right)|-\beta,-\alpha\rangle\right\},\nonumber\\
&&\hspace{-5mm}\psi_{e}(t)=N_{e}\left\{e^{-i\frac{\varphi}{2}}\cos\left(\frac{\varphi}{2}\right)\,
|\beta,\alpha\rangle
+ie^{i\frac{\varphi}{2}}\sin\left(\frac{\varphi}{2}\right)|-\beta,-\alpha\rangle\right\},
\end{eqnarray}
where we have used $N_{(g,e)}=\sqrt{2}$ and $gt=2k\pi$. Clearly, by
choosing $\varphi=2\pi$ and $\pi$, regardless of the atomic
detection, the state of the fields result the two variable coherent
states $|\beta,\alpha\rangle$ and $|-\beta,-\alpha\rangle$
respectively.
\subsection{Generation of the exited coherent states
$|\beta, \alpha; \mathbf{n}\rangle$} At this stage we follow an
approach that results the realization of the framework for the
production of the excited states $|\beta, \alpha;
\mathbf{n}\rangle$. These states can be produced by interaction of a
two level atom with a two mode cavity field $|\beta, \alpha\rangle$,
which can be described with the interaction Hamiltonian:
\begin{eqnarray}&&\hspace{-7mm}H_{int} = \hbar\mu\sigma_{+}a + h.c.\end{eqnarray}
Assume that, we are preparing the initial state of the system as
\begin{eqnarray}&&\hspace{-7mm}|\Psi(t= 0)\rangle = |\beta,
\alpha\rangle|e\rangle,\end{eqnarray} then, in the next and small
enough time we have
\begin{eqnarray}&&\hspace{-7mm}|\Psi(t)\rangle = e^{-\frac{i t}{\hbar}H_{int}}|\beta,
\alpha\rangle|e\rangle,\nonumber\\
&&\hspace{9mm}\vdots\nonumber\\&&\hspace{8mm}\Downarrow \nonumber\\
&&\hspace{-6mm}|\Psi(t)\rangle= |\beta, \alpha\rangle|e\rangle-{i\mu
t}a^{\dag}|\beta, \alpha\rangle|g\rangle.\end{eqnarray} Now,
depending on the situation in which the atom is detected, the state
of the two mode field may be determined. If the atom is detected in
a ground state, the state of the field will be collapsed to
$a^{\dag}|\beta, \alpha\rangle$. Finally, extension the above
arguments to multi photon process, the outcome photon field will be
a state $\left(a^{\dag}\right)^{n}|\beta, \alpha\rangle$, which are
the new type of the excited ( or photon added) coherent states
$|\beta, \alpha, \mathbf{n}\rangle$, were discussed above.
\section{Discussion and Outline }
Based on the action of the creation operator of the Weyl-Heisenberg
algebra, broad range of states that are called two variable
Agarwal's type of PACSs are produced , for first time based on our
knowledge. Realization of resolution of the identity condition, be
obtained with respect to the non-oscillating and positive definite
measure on the complex plane. Finally, non-classical properties of
such states have been reviewed in detail. For instance, it has been shown that:\\
$\bullet$ Their squeezing properties, which could be varied not only
by the magnitude of the external magnetic field $B_{ext}$, also by
the quantum number $\mathbf{n}$ and by the complex variables
$\beta$.
This allows to the experimenter to adjust the appropriate parameters to control the quantities.\\
$\bullet$ They result that, squeezing properties in $p_{x}$ is
efficiently considerable where $B_{ext}$ is decreased. However, we would not expect to take squeezing in $x$ component.\\
$\bullet$ One can show that $|\beta, \alpha, \mathbf{n}\rangle$
satisfies sub-Poissonian statistics, that is independent of the choice of the parameters.\\
$\bullet$ It is important to note that we will see the fully
squeezing in $|\beta, \alpha, \mathbf{n}=1\rangle$ and $|\beta,
\alpha, \mathbf{n}=0\rangle$, while the latter minimize the minimum uncertainty condition too.\\
$\bullet$ Finally it is worth to mention that the scheme proposed
here to construct PACSs in terms of the Glauber two variable
coherent states, could be applied to two other classes of minimum
uncertainty coherent states which were introduced in
\cite{dehghani0}.\\\\\textbf{Acknowledgments}\\
This work has been supported by a grant/research fund from
Azarbaijan Shahid Madani University.

\end{document}